\documentclass[12pt]{article}
\usepackage{graphicx}
\usepackage{caption}
\usepackage{amsmath}
\usepackage{bm}
\usepackage{setspace}
\usepackage{titlesec}
\usepackage{enumerate}
\usepackage{multirow}
\usepackage{etoolbox}
\usepackage[margin=1in]{geometry}
\usepackage{hyperref}
\usepackage{subfig}
\usepackage{float}

\begin{document}
	
\title{ZICS: an application for calculating the stationary probability distribution of stochastic reaction networks}
\author{\textbf{Michail Vlysidis$^{1}$, Andrew C. Schiek$^2$, Yiannis N. Kaznessis$^{3, 4*}$}\\
{\footnotesize Department of Chemical Engineering and Materials Science, }\\
{\footnotesize University of Minnesota, Minneapolis, MN 55455, USA}\\
{\footnotesize $^1$ vlysi001@umn.edu}
{\footnotesize $^2$ schie248@umn.edu}
{\footnotesize $^3$ yiannis@umn.edu}\\
{\footnotesize $^4$ General Probiotics Inc., St. Paul, MN 55114-1964, USA, yiannis@gprobiotics.com}\\
{\footnotesize $^*$ Corresponding author}}
\date{}
\maketitle

\begin{abstract}
	
Stochastic formalisms are necessary to describe the behavior of many biological systems. However, there remains a lack of numerical methods available to calculate the stationary probability distributions of stochastic reaction networks. We have previously development a numerical approach to calculate stationary probability distributions of stochastic networks, named ZI closure scheme. In this work, we present ZICS (Zero-Information Closure Scheme), a free applications based on ZI-closure scheme.

\end{abstract}

\section{Introduction}

Biochemical processes are governed by inherent thermal noise and stochasticity and many challenges remain in finding models to accurately describe their behavior. The chemical master equation (CME) is the most detailed mathematical model available, but its usage is intractable when applied to anything but the most simplistic systems \cite{Gillespie1992}. While kinetic Monte-Carlo algorithms (e.g SSA \cite{Gillespie1976}) have been developed to numerically solve biochemical reaction networks, the majority of these algorithms are connected to significant computational cost. An alternative approach, which we focus on, is to study the moments of the probability distribution \cite{Lakatos2015}.

There are numerous successful algorithms in the literature to generate moment equations \cite{Gillespie2009a,Constantino2016}. In this work we employ the method reported by Smadbeck and Kaznessis \cite{Smadbeck2012}, that can be applied for both polynomial and factorial moments. For nonlinear systems, lower-order moments depend on higher-order ones. As a result, moment equations are challenging to solve. Moment closure techniques emphasize on how to relate the higher and lower order moments. 

Smadbeck and Kaznessis \cite{Smadbeck2013} closed the scheme by relating all the moments to a set of Lagrange multipliers through maximizing the system's entropy. The suggested algorithm, named Zero-Information (ZI) Closure Scheme, is a computationally efficient way to calculate the stationary probability distribution of biochemical reaction networks. 

ZI closure scheme is an innovative and promising approach in the field of stochastic kinetics. However, the algorithmic implementation of the method can be computationally challenging to a non-expert. In order to make the method more broadly accessible and foster the progress in the field, we have translated the ZI closure scheme into a user friendly application. The application, named ZICS, is a Windows standalone program. The application is compatible with Matlab, version 2016b and later, so it can be used on non-Windows machines. It accurately solves multidimensional multistable non-linear reaction networks. ZICS in publicly available at GitHub (\href{https://github.com/mvlysidis/ZICS}{https://github.com/mvlysidis/ZICS}).

\section{Approach}

For a detailed description of the moment equation generator, we refer to \cite{Smadbeck2012}. In our previous work (\cite{Smadbeck2013} and \cite{Constantino2016}), we report a thorough analysis on the connection between moments and Lagrange multipliers ($ \boldsymbol{\lambda} $) through the maximization of entropy. 

In brief, for a reaction network with $N$ components, the stationary probability distribution $P(X_1 \ldots X_N)$ for a given composition $(X_1 \ldots X_N)$ is given by:
\begin{equation}
P(X_1 \ldots X_N)= \exp \left[ - \sum_{i=0}^{\Psi} \lambda _i f_{\mu _i} (X_1 \ldots X_N) \right]
\end{equation}
where $\lambda _i $ is the $i^{th}$ Lagrange multiplier and $f_{\mu _i} (X_1 \ldots X_N)$ represents the functional form of the lower-order moment $i$ evaluated at $(X_1 \ldots X_N)$. For example, if the third moment is the first combined factorial moment of $X_1$ and $X_2$, $\{X_1 \cdot X_2\}$, then $f_{\mu _3} (X_1 \ldots X_N) = X_1 \cdot X_2$. $\Psi$ is the number of lower-order moments. 

As a result, all the moments (both lower ($ \boldsymbol{\mu} $) and higher ($ \boldsymbol{\mu '} $) order) are related to Lagrange multipliers. The $i^{th}$ moment ($\mu _i$) is given by:
\begin{equation}
\mu _i = \sum_{\Omega} f_{\mu _i} (X_1 \ldots X_N) P(X_1 \ldots X_N) \nonumber
\end{equation}
\begin{equation}
= \sum_{\Omega} \left\{ f_{\mu _i} (X_1 \ldots X_N) \exp \left[ - \sum_{j=0}^{\Psi} \lambda _j f_{\mu _j} (X_1 \ldots X_N) \right] \right\}
\end{equation}
where $\Omega$ corresponds to the $N$-dimensional state space for all the possible values of $(X_1 \ldots X_N)$. The summation $\sum_{\Omega}$ represents $N$ multiple summation signs. 

It should be noted that the zero-order moment ($\mu _0$) is always 1, as $\sum_{\Omega} P(X_1 \ldots X_N) = 1$. Thus, the associated Lagrange multiplier $\lambda _0$ is not independent:

\begin{equation}
\lambda _0 = \log \left\{ \sum_{\Omega}  \exp \left[ - \sum_{i=1}^{\Psi} \lambda _i f_{\mu _i} (X_1 \ldots X_N) \right] \right\}
\end{equation}

We employ a Newton-Raphson algorithm that differs from the one reported by \cite{Smadbeck2013}. The residual is calculated as $R = A \boldsymbol{\mu} + A' \boldsymbol{\mu '} + \boldsymbol{\mu _c} $ based on the moment equations. Matrices $A$, $A'$ and the vector of constants $\boldsymbol{\mu _c}$ of the moment equations are generated as described in \cite{Smadbeck2012}. The Jacobian matrix is given by: $J_{i,j} = \frac{\partial \mu _i}{\partial \lambda _i} = - \mu _{i,j} + \mu _i \cdot \mu _j$, where $\mu _{i,j} = \sum_{\Omega} \left[ f_{\mu _i} (X_1 \ldots X_N)  f_{\mu _j} (X_1 \ldots X_N) P (X_1 \ldots X_N) \right]$ \cite{Vlysidis2018}. The algorithm of the application can be found in \cite{Vlysidis2017}.

\section{Methods}

ZICS was created with the Matlab app designer tool. The application runs on Windows machines independent of Matlab installation. To ensure this, the ZICS installation package includes Matlab runtime. For non-Windows machines, the application runs through Matlab, version 2016b and later.

ZICS calculates the stationary probability moments of a reaction network and reconstructs the probability distribution. It is also able to generate the factorial moment equations. The application assumes that each reaction has an elementary reaction rate. ZICS can only be applied to open systems, which we refer to networks with only the minimum number of independent components. In the section \ref{sec:mm}, we present an example how an non-open system can be transformed into a closed one. The user has to specify the reaction network, kinetic constants and the number of lower-order moments (closure order).

For the construction of the reaction networks, the number of reactions and components are necessary. The definition of reaction networks has the form of matrices for reactant and product stoichiometries as well as the kinetic constants of each reaction. After the creation of the network, the user indicates the Newton-Raphson parameters. The limits of the state space for each component are specified. The maximum closure order is also required. The application can terminate the calculations when a new order closure does not improve the solution significantly, before the maximum order of closure is reached. Finally, the user can change the initial guess of the Lagrange multipliers. The default guess is a uniform distribution. For more details about the functional components of the application, the interested user is directed to the manual located at the same folder as the source code.

\begin{figure}[H]
	\includegraphics[width=3in]{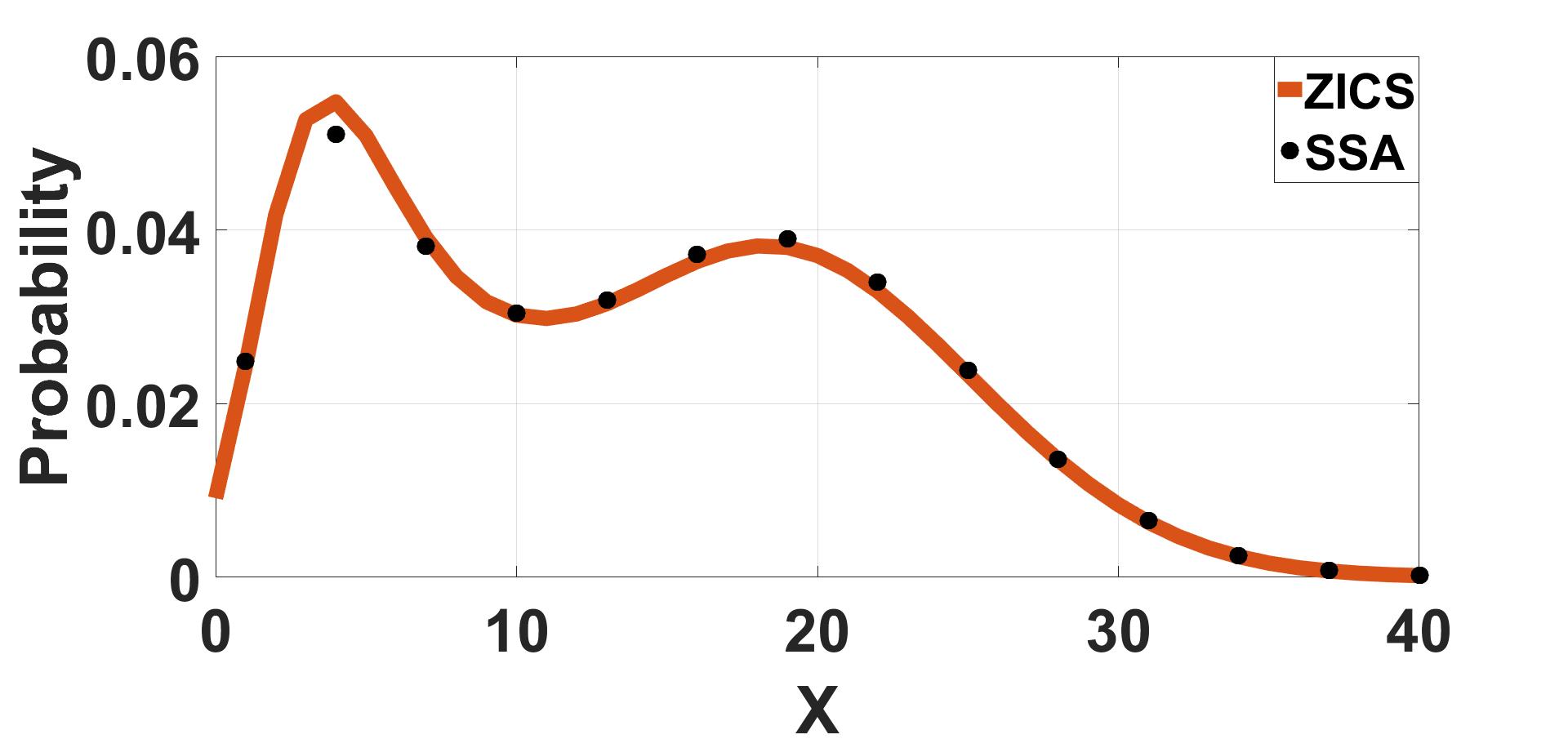}
	\includegraphics[width=3in]{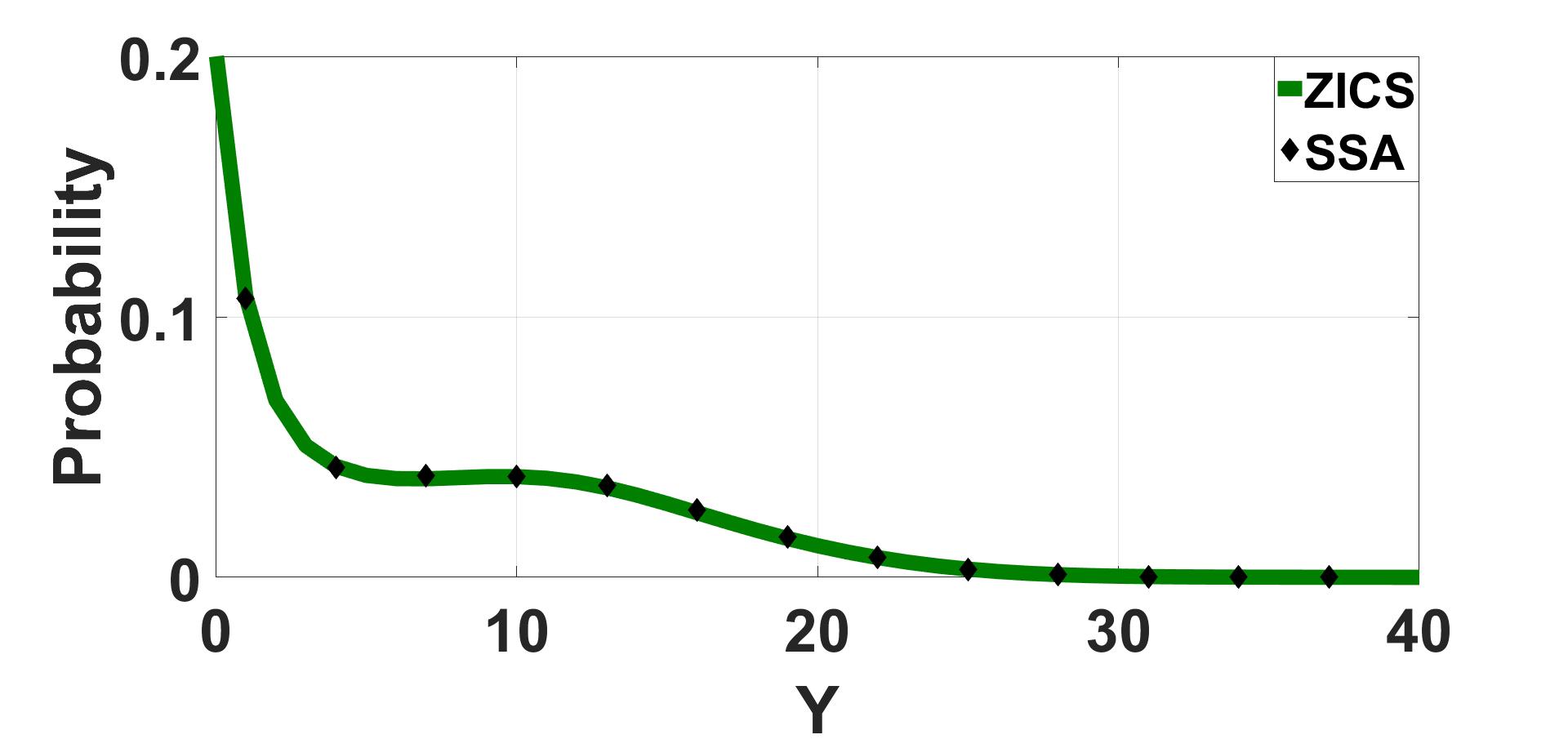}
	\caption[Stationary probability distribution of Wilhelm's network with ZICS application]{Stationary probability distribution of Wilhelm's network \cite{Wilhelm2009a} as obtained with ZICS application. Left figure presents component X and right figure component Y. The results of compared with SSA ones (solid dots). The kinetic constants are reported in Table \ref{table_system_app}.}
	\label{fig:01}
\end{figure}

Results for Wilhelm's network \cite{Wilhelm2009a} (Table \ref{table_system_app}), a two-component multistable non-linear network, are presented in figure \ref{fig:01}. Each component performs a different bistable behavior, that the application is able to capture at steady state. The applications results are as accurate as Gillespie's SSA \cite{Gillespie1976} ones. Graphs can be customized through the application. At the source code folder on GitHub, solutions for more than ten non-linear networks with ZICS are included (\href{https://github.com/mvlysidis/ZICS}{https://github.com/mvlysidis/ZICS}). 

\newpage

\section{Manual}

This section presents the main features of ZICS application with the goal to educate the reader on how to use the application. In order to demonstrate the application's aspects and capabilities, Wilhelm's stochastic reaction network \cite{Wilhelm2009a} is employed (Table \ref{table_system_app}).

\begin{table}[H]
	\centering
	\caption[Wilhelm's reaction netowrk]{The table shows the reactions for Wilhelm's bistable model and its kinetic constants. The network is used as an example to explain the application components.}
	\label{table_system_app}
	\begin{tabular}{cc}
		\hline
		\textit{\textbf{Wilhelm's Reaction Network}} & \textbf{Kinetic constants} \\ \hline
		$ Y \stackrel{k_1}{\longrightarrow}  2 X $ & $ k_1 = 35 $ \\
		$ 2 X \stackrel{k_2}{\longrightarrow}  X + Y $ & $k_2 = 1$ \\ 
		$ X + Y \stackrel{k_3}{\longrightarrow}  Y $ & $k_3 = 1$ \\
		$ X \stackrel{k_4}{\longrightarrow} \emptyset $ & $k_4 = 9.74$ \\ 
		$\emptyset \stackrel{k_5}{\longrightarrow} X $ & $k_5 = 30$ \\ \hline
	\end{tabular}
\end{table}

\subsection{Input the Network}
\label{sec1}

\begin{figure}[H]
	\centering
	\includegraphics[width=4in]{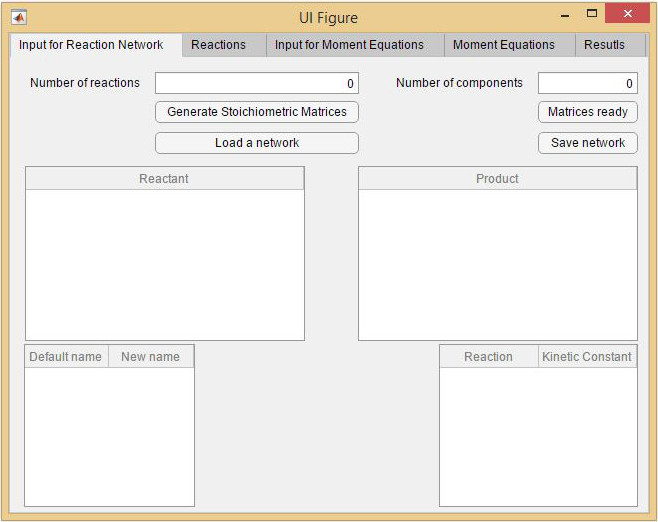}
	\caption[First tab of the ZICS application]{\label{fig1} First tab of the ZICS app.}
\end{figure}

In order to start solving the system, the user needs to input the reaction network and the kinetic constants into the application. The network is represented by the stoichiometric matrices for the reactants and products. Each reaction should be an irreversible reaction. Figure \ref{fig1} presents the starting tab ("Input for Reaction Network") of the application. 

\begin{figure}[H]
	\centering
	\subfloat[]{\includegraphics[width=2.75in]{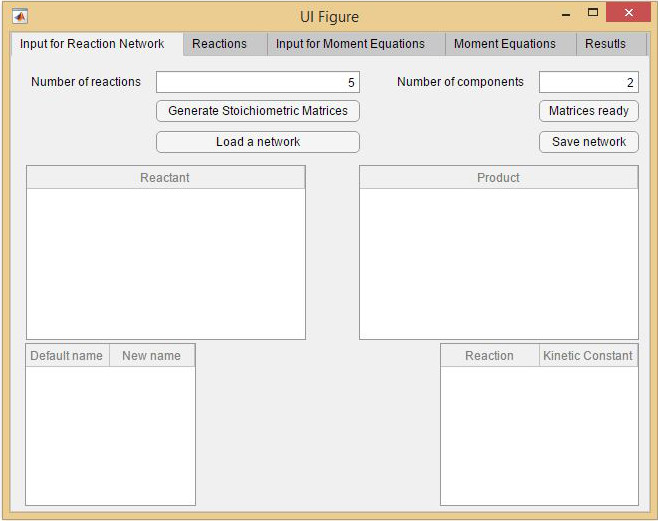}\label{fig2}}
	\subfloat[]{\includegraphics[width=2.75in]{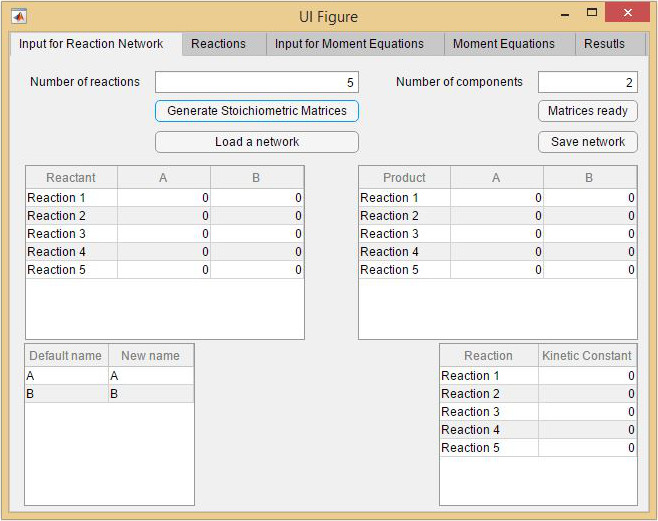}\label{fig3}}
	\caption[Number of reactions and components in ZICS]{(a) The number of reactions (5) denoted at top left box and the number of components (2) at the top right box. (b) The application creates blank matrices for the network after the "Generate Stoichiometric Matrices" button is pressed.}
\end{figure}

The number of reactions and components are the first options to be defined. Each reaction represents an irreversible reaction, thus a reversible reaction is represented as two separate ones. The number of components includes both the reactants and products. For the network at Table \ref{table_system_app}, the number of reactions is 5 and components 2 ($X$ and $Y$). The user can insert the number of reaction in the top left box and the number of components in the top right (figure \ref{fig2}).

After the number of components and reactions have been specialized, the user can create the reaction network. The user should press the "Generate Stoichiometric Matrices" button and the application creates blank matrices for the reactants, products, name of components and kinetic constants (figure \ref{fig3}). 

\begin{figure}[H]
	\centering
	\includegraphics[width=4in]{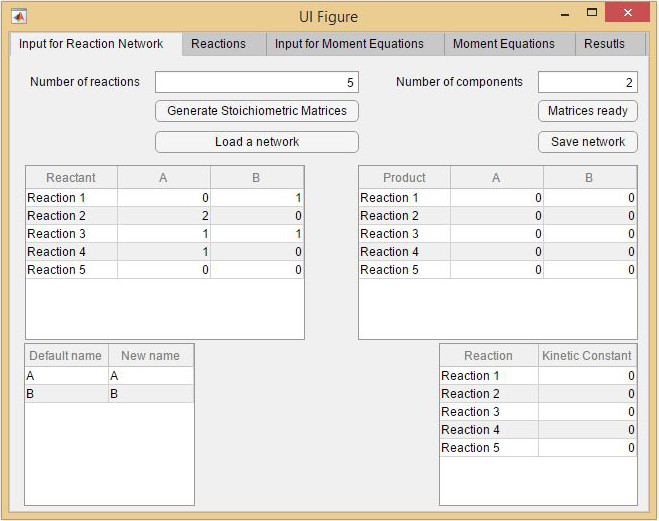}
	\caption[Fill the reactants stoichiometric matrices with ZICS ]{\label{fig4} The reactants stoichiometric table (top left) is filled. Each row represents a reaction and each column a component. The table is filled according the reactions of Table \ref{table_system_app}. Column A is used for component $X$ and column B for component $Y$.}
\end{figure}

The user can now start inputing the reaction network. The user can first denote the stoichiometric matrix for the reactants in the top left table with the "Reactant" indicator. The columns of the table denote the different components and the row the reaction of the network. In this case, we have two columns for the two components and five rows for the five reactions. We'll use column A for component $X$ and column B for component $Y$. From Table \ref{table_system_app}, it is easy to see that there are no molecules of $X$ and 1 molecule of $Y$ involved in the left side of the reaction. Thus, the first row-first column box has 0 and the first row-second column has 1. At the second reaction, 2 molecules of $X$ and none of $Y$ react. So the second row-first column box has 2. Following a similar procedure the user can fill the rest of the reactants table (figure \ref{fig4}).

\begin{figure}[H]
	\centering
	\includegraphics[width=4in]{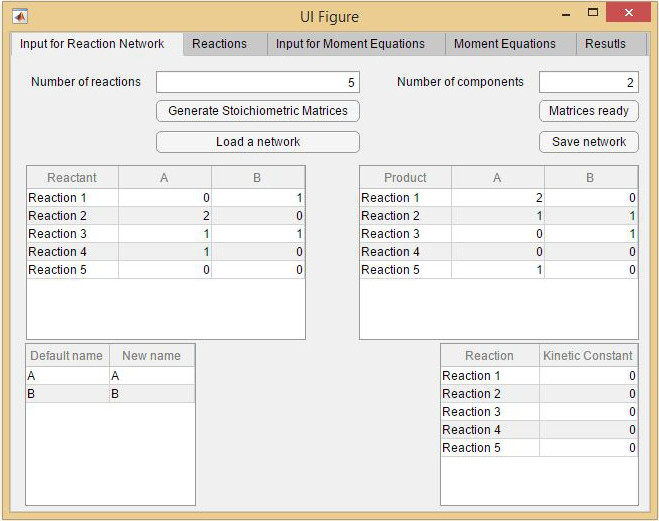}
	\caption[Fill the products stoichiometric matrices with ZICS ]{\label{fig5} The products stoichiometric table (top right) is filled. Each row represents the same reaction and each column the same component as in the reactants stoichiometric matrix. The table is filled according the reactions of Table \ref{table_system_app}.}
\end{figure}

Similarly, the user can input the stoichiometric matrix for the products in the top right table with the "Product" indicator. Each row of the table represents a reaction. The order of the reactions does not affect the solution, however each row of the reactants and products matrices should represent the same reaction. The columns of the table represent the components. In this case, column A refers to component $X$ and column B to component $Y$. Again, the pairing of columns and components does not influence the final result, however each column of the reactants and products matrices should represent the same component. For the reaction network presented at Table \ref{table_system_app}, the first reaction produces two molecules of $X$ and zero molecules of $Y$. Thus, the first row-first column box is filled with 2 and the first row-second column with 0. The second reaction produces one molecule of $X$ and $Y$, hence the second row-first column box is filled with 1 (for $X$) and the second row-second column box has also 1 (for $Y$). With the same strategy, the rest of the products stoichiometric matrix can be filled (figure \ref{fig5}). At this point the reaction network has been fully defined and the user can input the reaction kinetic constants.

\begin{figure}[H]
	\centering
	\includegraphics[width=4in]{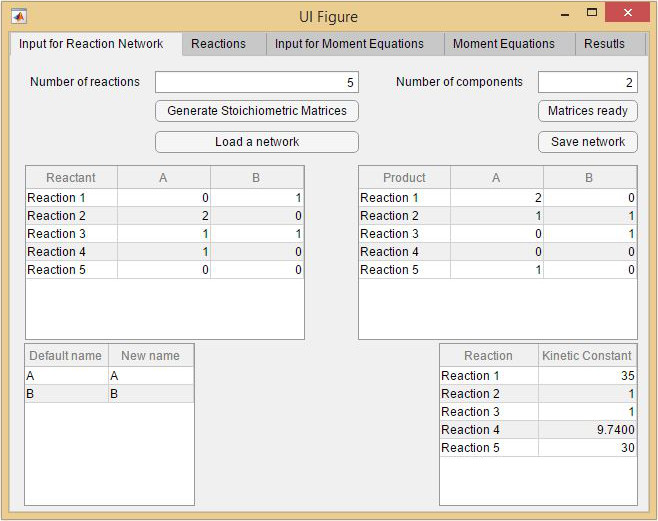}
	\caption[Reaction kinetic constatns with ZICS ]{\label{fig6} The kinetic constants table (bottom right) has been filled. Each row represents the same reaction as in the stoichiometric tables for the reactants and products. The value of the kinetic constants is based on Table \ref{table_system_app}.}
\end{figure}

The reaction constants can be inputted in the bottom right table indicated with "Kinetic Constant". The table has one row representing the value of the kinetic constants and multiple rows, one for each reaction. Each row should represent the same reaction as in the stoichiometric matrices for products and reactants. Based on Table \ref{table_system_app}, the first element of the table should 35 (for the kinetic constant of the first reaction), the second 1, the third 1 etc. (figure \ref{fig6}).

\begin{figure}[H]
	\centering
	\includegraphics[width=4in]{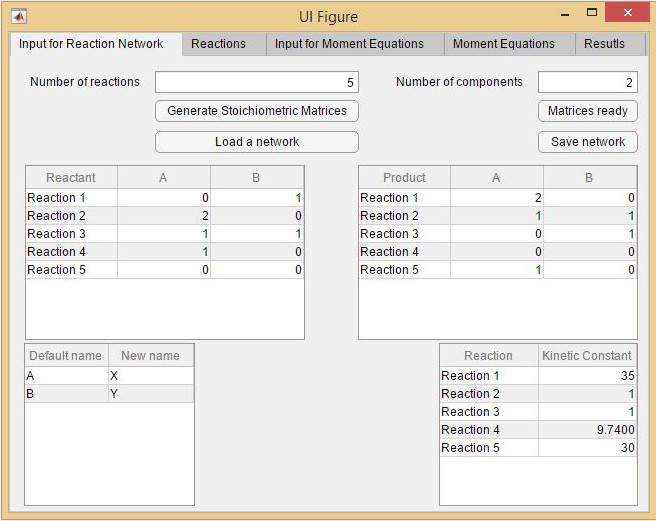}
	\caption[Rename components on ZICS]{\label{fig7} At the bottom left table, the user can rename the components. A and B represent $X$ and $Y$, respectively.}
\end{figure}

Finally, the application gives the ability to the user to rename the reactions components. The default option is the letters of the alphabet. The user can rename the components at the bottom left table. In this example, component A represents $X$ so it is renamed as $X$ and component B as $Y$, as shown in figure \ref{fig7}.

The user has also the ability to save the network. The "Save network" button saves the network in a Matlab .mat format. The application can load the previously saved networks with the "Load a network" button. This function allows to load sbml files (.xml) in a .tsv format. In order to first transform the .xml files to .tsv, the user is directed to free available website: \url{https://rumo.biologie.hu-berlin.de/SBtab/default/converter}.

\begin{figure}[H]
	\centering
	\includegraphics[width=4in]{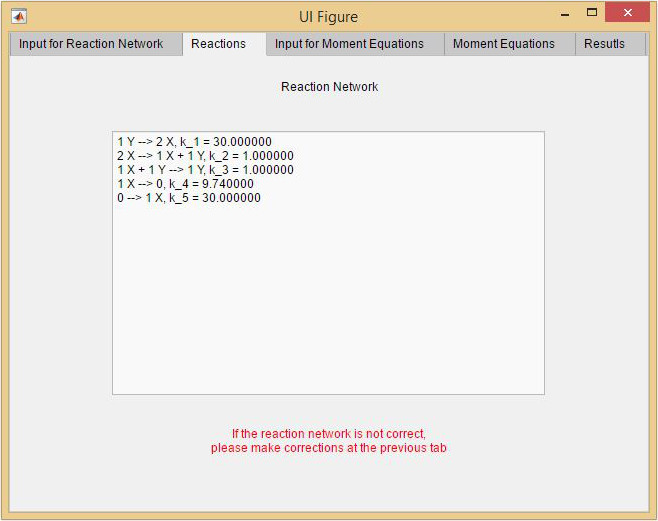}
	\caption[Second tab of ZICS]{\label{fig8} At the second tab named "Reactions" the user can check the form of the network. In case, an error has occurred the user can go back to the previous tab "Input for Reaction Network" and make the appropriate changes.}
\end{figure}

After the network, the kinetic constants and the name of the reactants have been inputted, the user can move on by pressing "Matrices Ready" button. The user has the ability to check the form of the network at the second tab ("Reactions"), as shown figure \ref{fig8}. If there is an error in the network, the user can go back to the previous tab ("Input for Reaction Network") and make the appropriate changes as described above. If changes in the "Input for Reaction Network" tab has been made, the user should press the "Matrices Ready" button to make them final and apply them for the rest of the application.

\subsection{Simulation Parameters}

\begin{figure}[H]
	\centering
	\includegraphics[width=4in]{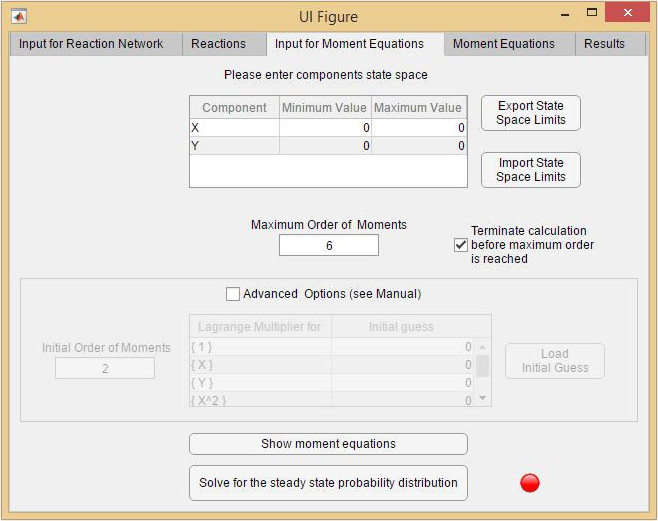}
	\caption[Third tab of ZICS]{\label{fig9} The simulation parameters can be entered at the "Input for Moment Equations" tab.}
\end{figure}

After the input of the reaction network and pressing the "Matrices ready" button, the user can enter the simulation parameters in order to solve the system. Simulation parameters can be entered at the third tab (named "Input for Moment Equations" as shown in figure \ref{fig9}). More details about the code of the application can be found in \cite{Smadbeck2013,Vlysidis2017}. 

The user can first specify the state space of the system. The state space of the system indicates the minimum and maximum number of molecules for each component. ZICS application solves stochastic reaction networks numerically and thus the state space of system should be indicated by the user and cannot be infinite. In the application, the state space is represented as a table (figure \ref{fig9}). Each row represents each of the components. There are two columns, one for the minimum value of the state space and one for the maximum value.

\begin{figure}[H]
	\centering
	\includegraphics[width=4in]{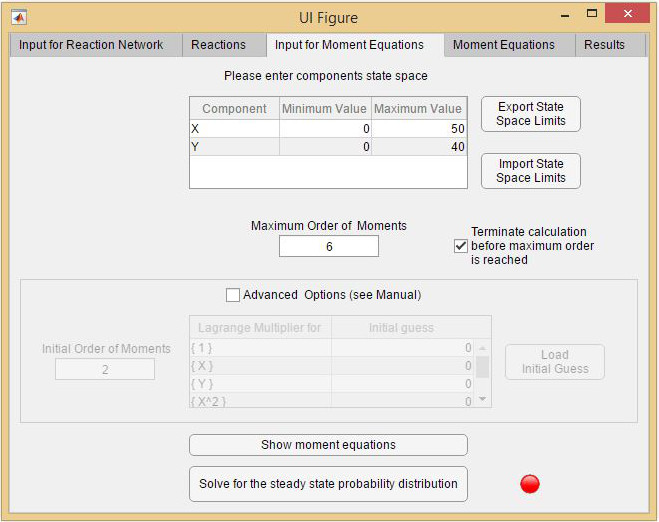}
	\caption[Specify network's state space in ZICS]{\label{fig10} The state space for component $X$ is between 0 and 50 and for component $Y$ is from 0 to 40.}
\end{figure}

It is suggested that the minimum value of each component is 0. The maximum value can vary per system and component. If the maximum value is less than the actual value, the application will produce inaccurate results, since it was not allowed to perform calculation in the whole necessary space. It is suggested that the maximum value is slightly higher than the actual value so that the application can perform calculations at the whole state space. There are two common approaches in case the user does not have a good educated guess about the maximum value. The first is to input an arbitrarily high numerical value. This can ensure accurate results, however the computational time required to produce results increases with the values of the state space. Thus, this approach can be computationally costly. An alternative way is to solve the system with a relatively average numerical value and then solve the the system with a higher maximum value and compare the two solutions. The final correct solution of the network should be independent of the state space values. For this example, we chose to use 50 molecules as the maximum value for component $X$ and 40 for component $Y$ as shown in figure \ref{fig10}. The user has the ability to save the state in a Matlab .mat format by pressing the "Export State Space Limits" button and load them again by pressing the "Import State Space Limits" button.  

\begin{figure}[H]
	\centering
	\includegraphics[width=4in]{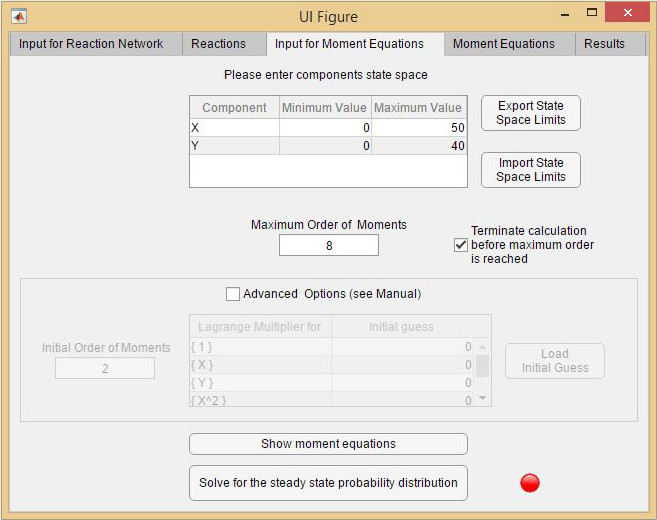}
	\caption[Maximum closure order in ZICS]{\label{fig11} The maximum order of moments can be modified. The application has the ability to terminate calculations before this number is reached. In this case, the maximum order of moments was changed to 8.}
\end{figure}

After the state space is specified, the user can enter the maximum order of moments. The application solves moment equations and the number of moments can change the size of the equations solved and thus the accuracy of the method. The maximum order of moments can be entered at the homonymous box (figure \ref{fig9}). For the majority of networks tested, an order of 6 moments is enough. The application has the ability to dynamically change the number of moments and test the accuracy of the results. Thus, the application can terminate the calculations before the maximum number of moments is reached if the solution meets the desired accuracy. This mode is automatically enabled. If desired, the user can disable this feature and force the application to perform calculations until the maximum number of moments is reached, by unselecting the box indicated as "Terminate calculation before maximum order is reached". It is suggested that the user inputs a high number of maximum moments and keeps the feature enabled. For this example, we chose that 8 order of moments is large enough (figure \ref{fig11}).

\begin{figure}[H]
	\centering
	\includegraphics[width=4in]{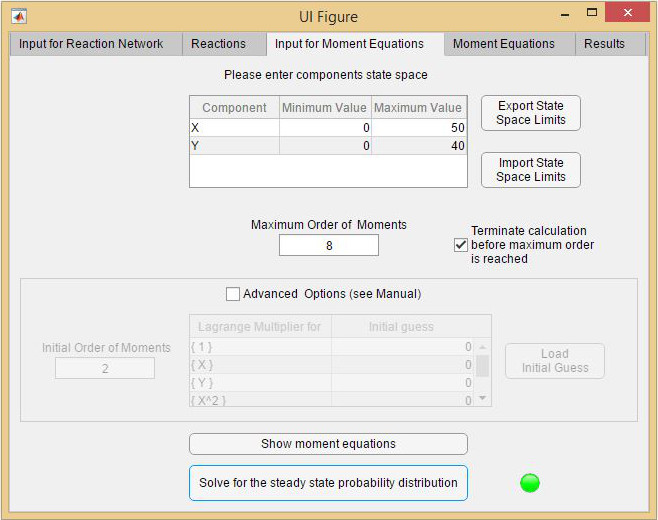}
	\caption[Produce results in ZICS]{\label{fig12} Calculations start when the "Solve for the steady state probability distribution" button is pressed. When the lamp next to the button is green, the calculations have finished.}
\end{figure}

With the order of moments and state space specified, the application is ready to produce results. The application calculates the stationary probability distribution for each component and their stationary probability moments. In order to start the calculations, the user should press the "Solve for the steady state probability distribution" button. Every time changes are performed in the "Input for Moment Equations" tab, the "Solve for the steady state probability distribution" button should be pressed to apply those changes. There is a lamp next to this button that indicates the status of the calculations. Before or during the calculations, the light of the lamp is red (figure \ref{fig11}). The light becomes green when the program has finished running and the stationary probability distribution for all the components of the network have been calculated (figure \ref{fig12}). 

\subsubsection{Advanced Options}
\label{sec2_1}

ZICS application is using ZI-closure scheme's Newton-Raphson algorithm. As such, the program requires an initial guess; the initial guess comes in the from of the initial number of moments and their associated Lagrange multipliers. The default initial condition is a uniform distribution with second order of moments and zero for all Lagrange multiplier values. The user has the ability to change the initial guess of the program by checking the "Advanced options (see Manual)" box. In the advanced options section, the user can modify the initial order of moments, the initial guess for Lagrange multipliers and also load an initial guess (by pressing the "Load Initial Guess" button). It is encouraged this section to mainly be used by experienced users.

\begin{figure}[H]
	\centering
	\includegraphics[width=4in]{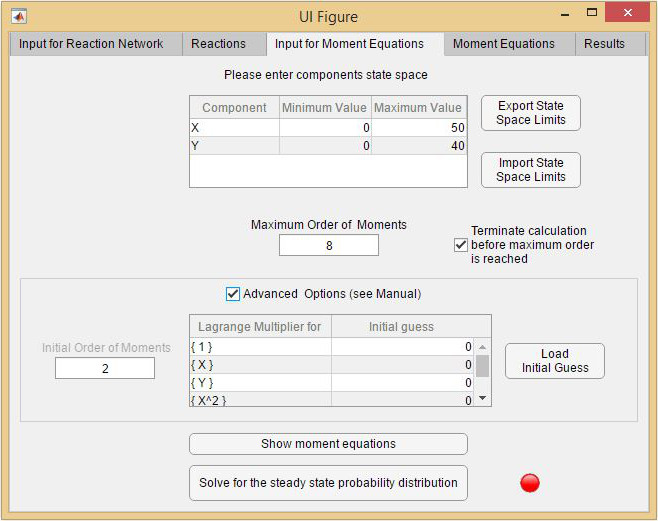}
	\caption[Advanced options]{\label{figadvanced} The user can input an initial guess for the algorithm by checking the "Advanced options (see Manual)" box.}
\end{figure}

\subsection{Results}
\label{sec3}

\subsubsection{Stationary Probability Distribution}

\begin{figure}[H]
	\centering
	\includegraphics[width=4in]{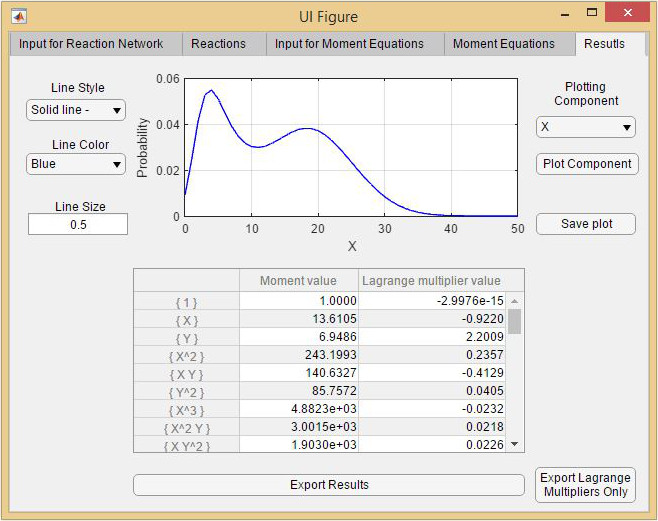}
	\caption[Fifth tab of ZICS]{\label{fig14} At the last tab named "Results" the user can find the solution of the network. The stationary probability distribution of the first component is automatically plotted. Additionally, the stationary moments and Lagrange multipliers are reported.}
\end{figure}

The results of the application are displayed at the last tab named "Results" (figure \ref{fig14}). The stationary probability distribution for the first component is automatically plotted at the top center. The x-axis represents the number of molecules of the component and the y-axis the probability. The plot can be exported by pressing the "Save plot" Button in multiple formats (.fig, .jpg, .png, .eps). Below the plot, the stationary values of the moments as well as their Lagrange multipliers are presented. The results can be saved in a Matlab (.mat) or spreadsheet (.xlsx) format by clicking the "Export Results" button. There is also the option to save the stationary Lagrange multipliers in a Matlab (.mat) format. Their values can be used at the advanced options section (section \ref{sec2_1}) for future runs. 

\begin{figure}[H]
	\centering
	\includegraphics[width=4in]{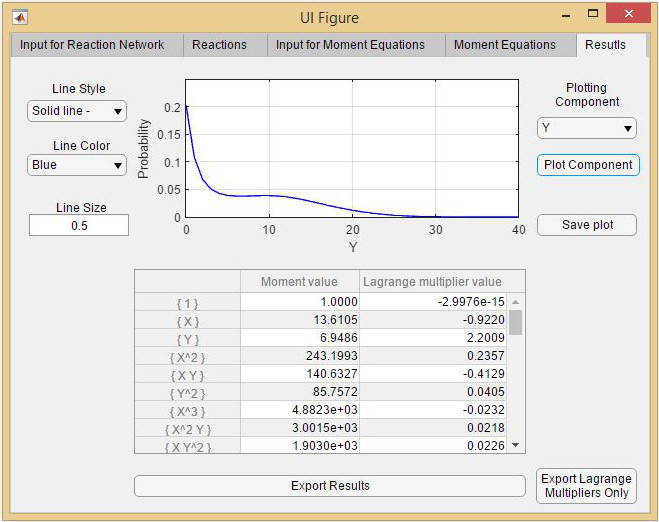}
	\caption[Plot diffrent components with ZICS]{\label{fig15} The "Plotting Component" drop-down menu gives the ability to plot different components. Here, the stationary probability distribution for component $Y$ is plotted after the press of "Plot Component" button.}
\end{figure}

\begin{figure}[H]
	\centering
	\includegraphics[width=4in]{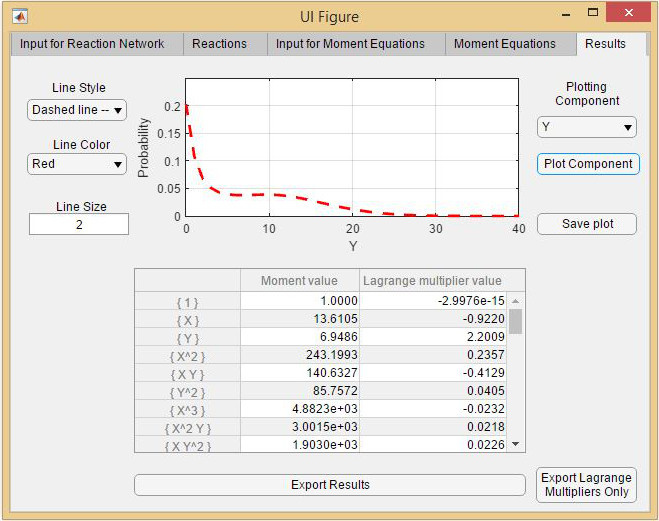}
	\caption[Modify plots]{\label{fig16} The user can modify the plot with the options of the left upper side and press the "Plot Component" button. Here, the "Line Style" is chosen to be dashed, the "Line Color" red and the "Line Size" 2.}
\end{figure}

The user has the ability to plot the stationary probability distribution for all the components of the system. By clicking the "Plotting Component" drop-down menu, the user can select the desired component. In order to plot the new component, the "Plot Component" button should be pressed (figure \ref{fig15}). The user can also choose between four different line styles, seven colors and change the line size of the plot from the respective options at the left-hand side of the plot (figure \ref{fig16}). Again, to apply any changes the user should press the "Plot Component" button. Any plots can be saved in four different formats (.fig, .jpg, .png or .eps) by pressing the "Save plot" Button. 

\subsubsection{Moment Equations}

\begin{figure}[H]
	\centering
	\includegraphics[width=6in]{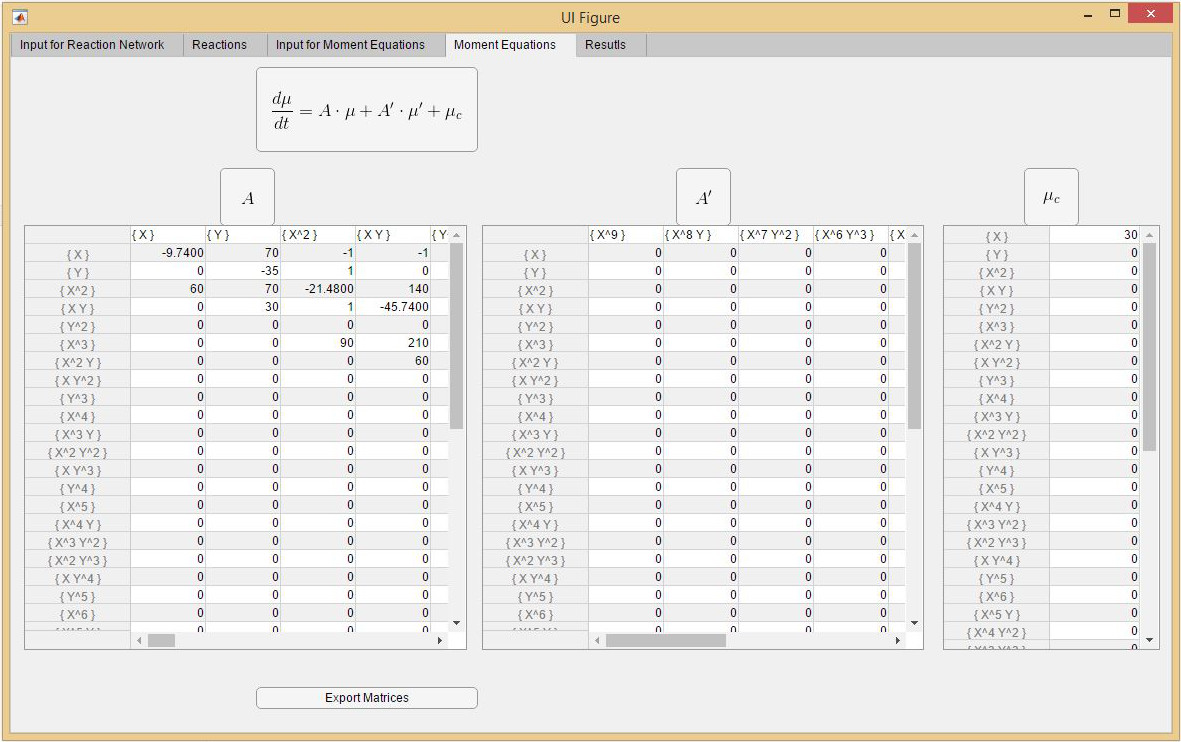}
	\caption[Fourth tab of ZICS]{\label{fig13} Moment equations matrices are displayed at the "Moment Equations" tab. This function is possible by pressing the "Show moment equations" button at the "Input Moment Equations" tab and independent of the function that calculated the stationary probability distribution.}
\end{figure}

Aside from calculating the stationary probability distribution of stochastic networks, the application also calculates the moment equations and its associated matrices. The moment equations are displayed at the "Moment Equations" tab, as shown in figure \ref{fig13}. In order for the matrices to appear, the "Show moment equations" button should be pressed at the "Input Moment Equations" tab (figure \ref{fig12}). This function is independent of the function that calculates the stationary probability distributions. The user can calculate the moment equation without solving the system. The moment equations can be exported in a Matlab (.mat) or spreadsheet (.xlsx) format by clicking the "Export Matrices" button.

\section{The Michaelis-Menten Reaction Network}
\label{sec:mm}

ZICS application is able to only solve open systems, which are systems with only the minimum possible number of components. Any closed system needs to be transformed into an open one in order to be solved with the application. In this section, the Michaelis-Menten reaction network \cite{Alberts2004,Chen2010} is used as an example of transforming a closed system to an open one. The closed form of Michaelis-Menten can be found in Table \ref{table_system_mm_app}.

The network has 2 degrees of freedom (components $S$ and $E$), however they are 4 components present ($S$, $E$, $S:E$ and $P$). The reaction network needs to be transformed into an open system based on the two relations: $E_{total}=E_T=S:E+E$ and $S_{total}=S_T=S:E+S+P$. Thus, components $S:E$ and $P$ depend on components $S$ and $E$ through $S:E = E_T - E$ \& $P = S_T - S - S:E = S_T - S - E_T + E$. Each reaction that includes either component $S:E$ or $P$ has to be reconstructed since these components are now dependent, and only independent components should remain in the reaction network. 

If the dependent components are products of a reaction, they can simply be removed from the reaction with no loss of information. For example, reaction 1 of the closed system of Table \ref{table_system_mm_app} (left column) is transformed from $ S + E \stackrel{k_1}{\longrightarrow}  S : E$ to $ S + E \stackrel{k_1}{\longrightarrow} \emptyset $ with the same kinetic constant. 

\begin{table}
	\centering
	\caption[Michaelis-Menten network in open and closed network format]{Michaelis-Menten network. Left columns show the networks closed form and the right ones show the open system.}
	\label{table_system_mm_app}
	\begin{tabular}{cc||cc}
		\hline 
		\multicolumn{2}{c||}{\textbf{Closed System}} & \multicolumn{2}{c}{\textbf{Open System}}  \\
		\textit{Reactions} & \textit{Kinetic Constants} & \textit{Reactions} & \textit{Kinetic Constants} \\ \hline
		\begin{tabular}{c}
			$ S + E {\longrightarrow}  S : E $  \\ 	$ S : E {\longrightarrow} S + E $  \\
			$ S : E {\longrightarrow} P + E $  \\	
		\end{tabular} &
		\begin{tabular}{c}
			$ k_1 $ \\	$ k_2 $ \\ $ k_3 $ \\ 
		\end{tabular} &
		\begin{tabular}{c} 	
			$S + E {\longrightarrow}  \emptyset$  \\
			$\emptyset {\longrightarrow} S + E$  \\
			$E {\longrightarrow} S + 2 E$  \\
			$\emptyset {\longrightarrow} E$  \\
			$E {\longrightarrow} 2 E$  \\ 
		\end{tabular} &
		\begin{tabular}{c} 	
			$k_1 $ \\ $E_T \ k_2 $  \\ $- k_2 $ \\
			$E_T \ k_3 $ \\ $- k_3 $ \\ 
		\end{tabular} \\ \hline
	\end{tabular}
\end{table}

If the dependent components are reactants of a reaction, then the transformation of the reaction is as follows:

Reaction 2 of left column of Table \ref{table_system_mm_app} is:
\begin{equation*}
S : E \stackrel{k_2}{\longrightarrow} S + E
\end{equation*}
The reaction can be transformed into:
\begin{equation*}
\emptyset \stackrel{k_2 \cdot S:E}{\longrightarrow} S + E
\end{equation*}
with a new kinetic constant of $k_2 \cdot S:E$. $S:E$ can be now substituted with $S:E = E_T - E$ from the mass balances and hence the reaction is: 
\begin{equation*}
\emptyset \stackrel{k_2 \cdot (E_T - E)}{\longrightarrow} S + E
\end{equation*}
with a new kinetic constant of $k_2 \cdot (E_T - E) = k_2 \cdot E_T - k_2 \cdot E$.  Since, the kinetic constant has two parts ($k_2 \cdot E_T$ and $- k_2 \cdot E$), the reaction can be split into two reactions:
\begin{equation*}
\emptyset \stackrel{k_2 \cdot E_T}{\longrightarrow} S + E \qquad
\emptyset \stackrel{- k_2 \cdot E}{\longrightarrow} S + E
\end{equation*}
Now the reaction includes only independent components, however the kinetic constant does not have a constant value any more with the presence of component $E$. The user needs to remove component $E$ from the kinetic constant and input it in the reaction as both a reactant and a product:
\begin{equation*}
\emptyset \stackrel{k_2 \cdot E_T}{\longrightarrow} S + E \qquad
E \stackrel{- k_2}{\longrightarrow} S + 2 E
\end{equation*}
Now, closed reaction 2 has been transformed into two open reactions. With the same methodology, the reconstructed open Michaelis-Menten reaction network is displayed on Table \ref{table_system_mm_app}.

\section{Summary}

We present ZICS, a standalone Windows application, for the calculation of the stationary probability distribution of stochastic biochemical reaction networks. With minimal computational cost, it accurately solves multistable multidimensional networks. ZICS also offers the ability to generate the factorial moment equations of a given network. ZICS is designed with the Matlab App deisgner. For non-Windows machines, the application can be used through Matlab, version 2016b and later.

The application is open source and publicly available (\href{https://github.com/mvlysidis/ZICS}{github.com/mvlysidis/ZICS}). At the source code folder, there are more than ten reaction networks that the application accurately solves. The networks vary in complexity, probability distribution modes and number of components, including the bistable Schl\"{o}gl model \cite{Schlogl1971}, the multistable Wilhelm's system \cite{Wilhelm2009a} and a five-component limit cycle. Most importantly, the application runs locally and requires an insignificant amount of computational time to produce accurate results.

\section*{Funding}

This work was supported by the National Institutes of Health [Grant  No.  GM111358]; National Science Foundation [Grant  No.  CBET-1412283]; Extreme  Science  and  Engineering  Discovery  Environment (XSEDE)  [NSF Grant  No.  ACI-10535753]; Minnesota  Supercomputing  Institute  (MSI);  and  the  University  of Minnesota  Digital  Technology Center.

\bibliographystyle{unsrt}
\bibliography{my_collection}

\end{document}